\documentclass[aps,showpacs,amsmath,preprint,amssymb,12pt]{revtex4}
\usepackage{graphicx}

\begin{document}
{~}
\vspace{3cm}

\title{Boundary Value Problem for Black Rings
\vspace{1cm}
}
\author{Yoshiyuki Morisawa ${}^{1}$, Shinya Tomizawa${}^2$ and Yukinori Yasui${}^2$}
\affiliation{ 
${}^1$ Faculty of Liberal Arts and Sciences, Osaka University of Economics and Law, Yao City, Osaka 581-8511, Japan\\
${}^2$Department of Mathematics and Physics,
Graduate School of Science, Osaka City University,
3-3-138 Sugimoto, Sumiyoshi, Osaka 558-8585, Japan
\vspace{3cm}
}

\begin{abstract}
We study the boundary value problem for asymptotically flat stationary black ring solutions to the five-dimensional vacuum Einstein equations. Assuming the existence of two additional commuting axial Killing vector fields and the horizon topology of $\rm \rm S^1\times \rm S^2$, we show that the only asymptotically flat black ring solution with a regular horizon is the Pomeransky-Sen'kov black ring solution. 
\end{abstract}

\preprint{OCU-PHYS 279}
\preprint{AP-GR 48}
\pacs{04.50.+h  04.70.Bw}
\date{\today}
\maketitle

\section{Introduction}

In four dimensions, stationary black hole spactimes have been studied by many authors and these studies are known as {\it uniqueness theorems of black holes}~\cite{uniqueness}. Israel showed that the only static, asymptotically flat vacuum solution with a regular event horizon is the Schwarzschild solution specified only by its mass~\cite{Israel}. Shortly afterward, he generalized the theorem to the situation with an electromagnetic field~\cite{Israel2}, i.e, he presented the proof that the only static, asymptotically flat electrovac solution with a regular non-degenerate event horizon is the Reissner-Nordstr\"{o}m solution with two parameters, the mass $m$ and the electric charge $q$ which are subject to the inequality $m^2>q^2$. Bunting and Masood-ul-Alam succeeded in proving these two theorems without the assumption of a single connected component of a black hole~\cite{Bunting,Bunting2} by using the positive energy theorem~\cite{Schoen,Witten}. On the other hand, for non-static and stationary  space-times, the Einstein-Maxwell system can be reduced to two-dimensional boundary value problem. The field equation is derived by the Ernst potential associated with the axial Killing vector field. The essential part in showing the uniqueness theorem for stationary black holes is that two solutions with the same asymptotic condition are isometric to each other. Carter~\cite{Carter} showed that the infinitesimally neighboring vacuum solutions with the same asymptotic conditions are equal, and Robinson generalized its theorem to the stationary electrovac system~\cite{Robinson}. Thereafter, using so-called {\it Robinson identity}, Robinson also succeeded in proving the uniqueness of the vacuum  Kerr family~\cite{Robinson2} with $m^2>a^2$ among all asymptotically flat, stationary and axisymmetric black hole solutions with a non-degenerate event horizon,
i.e., showing that two arbitrary, not necessarily infinitesimally neighboring, solutions with the same boundary conditions are equal to each other. In electromagnetic system, Mazur derived the divergence identity ({\it Mazur identity})~\cite{Mazur}, which is based on that the Ernst equation describe a non-linear sigma model on the symmetric space $SU(1,2)/S(U(1)\times U(2))$, and showed that the only possible and axisymmetric black hole solutions to the Einstein-Maxwell equations is the Kerr-Newman solution specified by the mass $m$, the angular momentum $j$ and the electric charge $q$ with the constraint $m^2>j^2/m^2+q^2$.

In recent years, studies of black holes in higher dimensions
have attracted much attention in the context of string theory
and the brane world scenario.
In fact, it has been predicted that higher-dimensional black holes
would be produced in a future linear collider
\cite{BHinCollider,kanti}.
Such physical phenomena are expected not only to give us
a piece of evidence for the existence of extra dimensions
but also to help us to draw some information toward quantum gravity.
Studies on stationary black hole solutions are important
since we may detect the Hawking radiation after the formation of stationary black holes in a collider.

A striking feature of asymptotically flat stationary black hole solutions in five dimensions is that they admit event horizons with non-spherical topologies in contrast to four dimensions. For instance, the topology of the event horizon
in higher dimensions cannot be uniquely determined
\cite{Cai,Helfgott,galloway}
in contrast to four-dimensional ones, which is restricted only to
the two sphere~\cite{Hawking,hawking_ellis}.
In five dimensions, however, the possible geometric types of the horizon topology are $\rm S^3$ and $\rm S^1\times \rm S^2$ \cite{Cai},
and in dimensions higher than five, more complicated
\cite{Helfgott,galloway}.
The black ring solutions with the horizon topology $\rm \rm S^1\times \rm S^2$, which rotate along the $\rm \rm S^1$ direction, were found by Emparan and Reall as solutions to the five-dimensional vacuum Einstein equations~\cite{Emparan:2001wn}. This is the first example of black hole solution with non-spherical topology. 
In addition to the black ring solution,
the rotating black hole solution with $\rm S^3$ horizon topology
had been already found by Myers and Perry \cite{Myers:1986un}.
Remarkably, within some range of the parameters, there are one black hole and two black rings with the same values of the mass and the angular momentum, which means the violation of the uniqueness known in four dimensions. Subsequently, other black ring solutions were found. The black ring solutions with a rotating two sphere were found by Mishima and Iguchi~\cite{MI}, and moreover, one with two angular momenta was constructed by Pomeransky and Sen'kov~\cite{Pom} by using the inverse scattering method~\cite{Belinskii, solitonbook,Tomizawa,Tomizawa2,Tomizawa3,Tomizawa4,Iguchiboost,Pomeransky:2005sj,Koikawa,Azuma,Iguchi,Castejon-Amenedo:1990b}.

For the asymptotically flat, static solutions
of higher-dimensional vacuum Einstein equations, the Schwarzschild-Tangherlini
solution \cite{T-schwarzschild} is the unique solution \cite{shiromizu},
and moreover, which is stable against linear perturbations
\cite{kodama}. It has been shown that the five-dimensional
Myers-Perry solution is unique if the topology is restricted to $\rm S^3$
and the spacetime admits three commuting Killing vectors
\cite{Morisawa:2004tc}. Hence it is natural to ask whether the Pomeransky-Sen'kov black ring solution is also unique under the assumptions of the existence of three commuting Killing vector field and the horizon topology of $\rm S^1\times S^2$. As mentioned above, however, there are two different black ring solutions for the same mass and the same angular momenta. Therefore, we must add some additional information to consider the bondary value problem for black ring solutions. One of the examples is the {\it rod structure} introduced by Harmark~\cite{Harmark}. By introducing the rod structure, Hollands and Yazadjiev~\cite{Hollands} applied the discussion of Morisawa and Ida to the case of non-spherical horizon topology and showed that two asymptotically flat and five-dimensional black hole solutions with the same topology, the same mass, the same angular momenta and the same rod structure are isometric to each other.

In this article, we study the boundary value problem for stationary black ring solutions to the five-dimensional vacuum Einstein equations. Assuming the existence of two additional commuting axial Killing vector fields and the horizon topology of $\rm \rm S^1\times \rm S^2$, we show that the only asymptotically flat black ring solution with a regular horizon is the Pomeransky-Sen'kov black ring solution. 
Our proof consists of two steps: First, we present a more general black ring solution than the Pomeransky-Sen'kov black ring solution in the sense that the solution, in general, has a conical singularity and is characterized by independent four parameters, i.e.,  the mass, two angular momenta and an additional parameter. By the requirement of the absence of a conical singularity on the solution, it coincides with the Pomeransky-Sen'kov black ring solution. Second, following the discussion in Ref.~\cite{Morisawa:2004tc,Hollands}, two arbitrary asymptotically flat black ring solutions with the same mass, the same two angular momenta and the same ratio of the radius of $\rm S^2$ to the radius of $\rm S^1$ as the solution are isometric. Hence we can conclude that the only asymptotically flat black ring solution without a conical singularity to the five-dimensional vacuum Einstein equations is the Pomeransky-Sen'kov black ring solution.

The remainder of this article is organized as follows: In Sec.\ref{sec:solution} we present general black ring solutions with four parameters. In Sec.\ref{sec:boundary}, we study the rod structure of the solutions. In Sec.\ref{sec:mazur}, we give a short explanation of the Mazur identity. In Sec.\ref{sec:coincidence}, using the Mazur identity, we show that an arbitrary asymptotically black ring solution with the same mass, the same two angular momenta and the same rod structure as our solution are isometric to it. In Sec.\ref{sec:conclusion}, we state the final results and the theorem.

\section{General Black Ring}\label{sec:solution}
\def\Ap{A^{+}_{xy}}
\def\Am{A^{-}_{xy}}
\def\Bpx{B^{+}_{x}}
\def\Bmx{B^{-}_{x}}
\def\Bpy{B^{+}_{y}}
\def\Bmy{B^{-}_{y}}
\def\Cpx{C^{+}_{x}}
\def\Cmx{C^{-}_{x}}
\def\Cpy{C^{+}_{y}}
\def\Cmy{C^{-}_{y}}
\def\Sx{S_{x}}
\def\Tx{T_{x}}
\def\Sy{S_{y}}
\def\Ty{T_{y}}
\def\SQ{Q}
\def\sq#1{s_{#1}}
\def\cY{Y}
\def\alphazero{\alpha}
\def\Dxy{D_{xy}}

The metric of general black ring solution, which in general has a conical 
singularity, is given by
\begin{eqnarray}
ds^2&=&-\frac{H(y,x)}{H(x,y)}(dt+\Omega)^2-\frac{F(x,y)}{H(y,x)}d\phi^2-2\frac{J(x,y)}{H(y,x)}d\phi d\psi+\frac{F(y,x)}{H(y,x)}d\psi^2\nonumber\\
    & &+\frac{2k^2H(x,y)}{(x-y)^2(1-\nu)^2}\left(\frac{dx^2}{G(x)}-\frac{dy^2}{G(y)}\right),
\label{eq:metric}
\end{eqnarray}
where the C-metric coordinates $x,y$ run the ranges of $-1\le x\le 1$ 
and $(-\lambda+\sqrt{\lambda^2-4\nu})/2\le y<\infty$ or 
$-\infty<y\le -1$, respectively. The solution has four independent parameters satisfying the 
inequalities $0\le \nu<1$, $2\sqrt{\nu}\le \lambda<1+\nu$, $k>0$ and $c\le b <1$ with
\begin{eqnarray}
c=\frac{\sqrt{\lambda^2-4\nu}}{1-\nu}.
\end{eqnarray}
The function $G$ appearing in the metric is defined as:
\begin{eqnarray}
G(x)&=&(1-x^2)(1+\lambda x+\nu x^2).
\end{eqnarray}
Since the other functions $H,J,F$ and the one-form $\Omega$ have considerably complicated forms, we do not write it here. The explicit expressions of them are given in Appendix~\ref{app:metric}. As will be mentioned later, under the choice of the parameters:
\begin{eqnarray}
b=\frac{2c}{1+c^2},
\end{eqnarray}
which is the condition for a conical singularity inside the black ring to vanish, the metric reduces to that of the Pomeransky-Sen'kov black ring solution.

\section{Boundary conditions}\label{sec:boundary}

\subsection{Rod Structure}
In this section, we give the rod structure of the black ring solution obtained in Sec.\ref{sec:solution}. To investigate the rod structure, we introduce the canonical coordinates defined by
\begin{eqnarray}
& & \rho^2=-\frac{4k^4G(x)G(y)}{(x-y)^4(1-\nu)^2},\\
& & z=\frac{k^2(1-xy)(2+(x+y)\lambda+2xy\nu)}{(x-y)^2(1-\nu)}.
\end{eqnarray}
Then, the metric can be written in the form
\begin{eqnarray}
ds^2=g_{\alpha\beta}dx^\alpha dx^\beta+\tilde f(d\rho^2+dz^2),
\end{eqnarray}
where $\alpha,\beta$ run $t,\phi,\psi$. The metric functions $g_{\alpha\beta}$ and $\tilde f$ depend only on the coordinates $\rho$ and $z$.

(i) The semi-infinite spacelike rod $[-\infty,-ck^2]$ and the finite rod $[ck^2,k^2]$ have the direction $v=(0,0,1)$, i.e., for $\rho=0$, $z\in[-\infty,-ck^2]$ and $\rho=0$, $z\in[ck^2,k^2]$, $g_{\alpha\beta}v^\beta=0$ holds. Since these give $g_{\psi\psi}=0$, $\rho=0$, $z\in[-\infty,-ck^2]$ and $\rho=0$, $z\in[ck^2,k^2]$ denote $\psi$-axis, i.e., the plane invariant under the rotation associated with the Killing vector $\partial_\psi$. For $\rho=0$ and $z\in[-\infty,-ck^2]$, the periodicity of the angular variable $\psi$ becomes
\begin{eqnarray}
\Delta\psi=\lim_{\rho\to 0}2\pi\sqrt{\frac{\rho^2 \tilde f}{g_{\alpha\beta}v^\alpha v^\beta}}=2\pi.\label{eq:period}
\end{eqnarray}
To cure conical singularities in the region $z\in[ck^2,k^2]$, one must impose the following condition on the parameters in the solutions:
\begin{eqnarray}
\Delta\psi&=&\lim_{\rho\to 0}2\pi\sqrt{\frac{\rho^2 \tilde f}{g_{\alpha\beta}v^\alpha v^\beta}}\\
          &=&2\pi \sqrt{\frac{-(-1+b)(-1+c)^2(c\alphazero+b(-2(1+c)^2+c\alphazero))^2}{(1+b)(1+c)^2(c\alphazero-b(2(-1+c)^2+c\alphazero))^2}},\label{eq:period2}
\end{eqnarray}
where the constant $\alpha$ is defined by
\begin{eqnarray}
\alphazero &=& 4 \frac 
{ ( \lambda+q ) ( \lambda+2-q ) ( \lambda-2-q ) }
{ ( \lambda-q ) ( \lambda+2+q ) ( \lambda-2+q ) }
\end{eqnarray}
with
\begin{eqnarray}
q &=& \sqrt{\lambda^2-4\nu}.
\end{eqnarray}
The periodicities (\ref{eq:period}) and (\ref{eq:period2}) of $\psi$ require putting the parameters as
\begin{eqnarray}
b=\frac{2c}{1+c^2},\label{eq:Pom_parameter}
\end{eqnarray}
\begin{eqnarray}
b=\pm\frac{c\alpha}{\sqrt{4-8c^2+4c^4+c^2\alpha^2}}.\label{eq:singular_parameter}
\end{eqnarray}
It should be noted that the solution with the parameters (\ref{eq:Pom_parameter}) exactly coincides with the solutions without a conical singularity obtained by Pomeransky and Sen'kov, which describe the rotating black ring  in two orthogonal planes independently, although the choice of the remainder parameters (\ref{eq:singular_parameter}) yields singular solutions.

(ii) The finite timelike rod $[-ck^2,ck^2]$ corresponds to an event horizon with topology $\rm \rm S^1\times \rm S^2$ since $\partial_\psi$ vanishes on both side of this rod. One see that $g_{\alpha\beta}v^\beta=0$ for $\rho=0$ and $z\in[-ck^2,ck^2]$. $v$ denotes the eigenvector with respect to the eigenvalue of zero and can be written in the form of $v=(1,\Omega_1,\Omega_2)$, where
\begin{eqnarray}
{{\Omega}_{1}}^{2}&=&
\frac{(1+b)(b-c)[2b(1-c)^2-(1-b)c\alphazero][2b(1-c^2)-(1-b)c\alphazero]}
{2(1-b)b(1-c)^2[2b(1+c)^2-(1+b)c\alphazero][2b(1-c^2)-(1+b)c\alphazero] k^2},
\\
{{\Omega}_{2}}^{2}&=&
\frac{(1+b)[2b(1-c^2)-(1-b)c\alphazero]}
{\alphazero(1-b)[2b(1-c^2)-(1+b)c\alphazero]}
\nonumber\\&\times&
\frac{(1+c)^2[2b(1-c^2)-c\alpha]^2[2b(1-c)^2-(1-b)c\alpha]^2}
{c^2[4b^2(1-c^2)^2-(1-b^2)c^2\alphazero^2]^2 k^2}.
\end{eqnarray}
Here the two constants $\Omega_1$ and $\Omega_2$ denote the angular velocities of the horizon along the directions $\partial_\phi$ and $\partial_\psi$, respectively.

(iii) The semi-infinite spacelike rod $[k^2,\infty]$ has the direction $v=(0,1,0)$.  Since these give $g_{\phi\phi}=0$, $\rho=0$, $z\in[k^2,\infty]$ means the $\phi$-axis, i.e., the plane under a rotation with respect to the Killing vector field $\partial_\phi$. The periodicity of the angular variable $\phi$ is computed as
\begin{eqnarray}
\Delta\phi=\lim_{\rho\to 0}2\pi\sqrt{\frac{\rho^2 \tilde f}{g_{\alpha\beta}v^\alpha v^\beta}}=2\pi.
\end{eqnarray}

\begin{figure}[htbp]
\begin{center}
\includegraphics[width=0.9\linewidth]{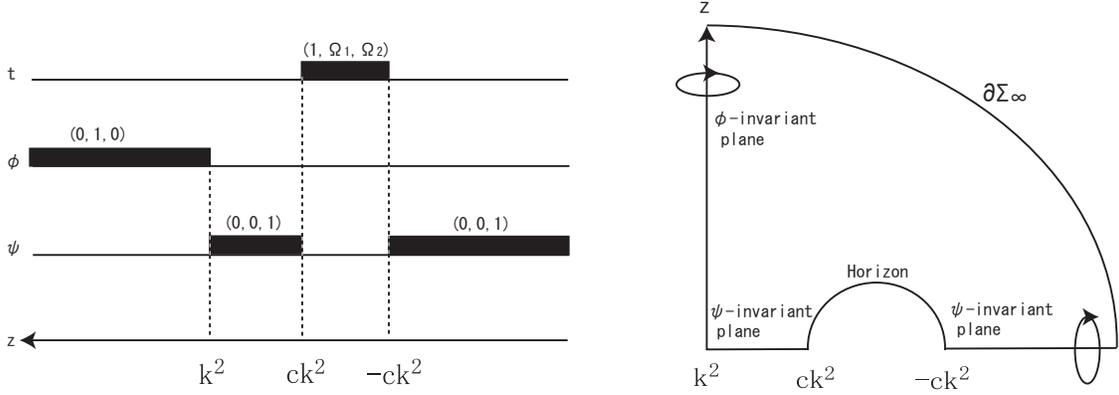}
\end{center}
\caption{The left figure show the rod structure of the five dimensional rotating black ring solution with asymptotic flatness. The vectors on the rods denote their directions. The right figure shows the black ring solution on the $(\rho,z)$-plane with the three-dimensional $(t,\phi,\psi)$ part suppressed.}
\end{figure}

\subsection{Asymptotic behavior}
Next, we introduce the coordinates $(r,\theta)$ defined by $\rho=\frac{r^2}{2}\sin2\theta$ and $z=\frac{r^2}{2}\cos2\theta$.
In the asymptotic region $r\to\infty$, the metric behaves as
\begin{eqnarray}
ds^2&\simeq& \left(-1+\frac{8M}{3\pi r^2}\right)dt^2-\frac{2J_\phi\sin^2\theta}{\pi r^2}dtd\phi-\frac{2J_\psi\cos^2\theta}{\pi r^2}dtd\phi\nonumber\\
    & &+dr^2+r^2(d\theta^2+\sin^2\theta d\phi+\cos\theta d\psi^2),
\end{eqnarray}
where the ADM mass $M$ and ADM angular momenta $J_\phi$, $J_\psi$ are given by 
\begin{eqnarray}
M=
\frac{3\pi b(1-c) [(1-b^2)c^2 \alphazero^2-4(1-c^2)^2b^2] k^2}
{2(1+b)[2b(1-c)^2-(1-b)c\alpha][2b(1-c^2)-(1-b)c\alpha]},
\end{eqnarray}
\begin{eqnarray}
{J_{\phi}}^{2}&=&
\frac{2\pi^2 (1-b)b(b-c)(1-c)^2[(1+b)c\alphazero-2b(1+c)^2]}
{[(1+b)c\alphazero-2b(1-c^2)]}
\nonumber\\&\times&
\frac{[(1+b)c\alphazero-2b(1-c)^2]^2 [(1-b^2)c^2\alphazero^2-4b^2(1-c^2)^2]^2 k^6}
{[(1-b)^2(1+b)c^2\alphazero^2-4b(1-b^2)(1-c)c\alphazero+4b^2(1+b)(1-c)^3(1+c)]^3},
\\
{J_{\psi}}^{2}&=&
\frac{16 \pi^2 b^2 (1-b) c^4 (1-c)^2 \alphazero}
{(1+b)[(1-b)c\alphazero-2b(1-c)^2]^2}
\nonumber\\&\times&
\frac{[(1-b^2) c^2 \alphazero^2-4 b^2 (1-c^2)^2]^2 k^6}
{[(1-b)c\alphazero-2b(1-c^2)]^3[(1+b)c\alphazero-2b(1-c^2)]}.
\end{eqnarray}

\section{Mazur Identity}\label{sec:mazur}
Here we give the brief review on the formalism developed in Ref.~\cite{Morisawa:2004tc}, where it is shown that the Myers-Perry solution is unique within a class of the five-dimensional asymptotically flat solutions with the horizon topology of $\rm S^3$ and additional two commuting spacelike Killing vectors. We consider the five-dimensional space-times admitting two commuting Killing vector fields $\xi_I=\partial_I (I=\phi,\psi)$. Then, the metric can be written in the form
\begin{eqnarray}
g=f^{-1}\gamma_{ij}dx^idx^j+f_{IJ}(dx^I+w^I_idx^i)(dx^J+w^J_jdx^j),
\end{eqnarray}
where $i,j$ run $1,2,3$ and $f={\rm det}(f_{IJ})$. The rescaled three-dimensional metric $\gamma_{ij}$, the metric functions $w^I_i$ and $f_{IJ}$ are independent of $\phi$ and $\psi$. The twist potentials $\omega_I$ are defined by
\begin{eqnarray}
d\omega_I=*(\xi_\phi\wedge\xi_\psi\wedge d\xi_I).
\end{eqnarray}
Then, the vacuum Einstein equations reduce to the system of the five scalar fields $f_{IJ}$ and $\omega_I$ on the three-dimensional space: 
\begin{eqnarray}
D^2f_{IJ}&=&f^{KL}Df_{IK}\cdot Df_{JL}-f^{-1}D\omega_I\cdot D\omega_J\label{eq:f},\\
D^2\omega_I&=&f^{-1}Df\cdot D\omega_I+f^{JK}Df_{IJ}\cdot D\omega_J,\label{eq:o}
\end{eqnarray}
and the Einstein equations for the three-dimensional space:
\begin{eqnarray}
{}^{(\gamma)}R_{ij}=\frac{1}{4}f^{-2}f_{,i}f_{,j}+\frac{1}{4}f^{IJ}f^{KL}f_{IK,i}f_{JL,j}+\frac{1}{2}f^{-1}f^{IJ}\omega_{I,i}\omega_{J,j},\label{eq:gamma}
\end{eqnarray}
where $D$ is the covariant derivative with respect to the metric $\gamma_{ij}$ and $\cdot$ denotes the inner product by $\gamma_{ij}$.
Here we assume the existence of another Killing vector field, a timelike Killing vector field $\xi_3=\partial_t$ which commutes with the other Killing vector fields $\xi_I(I=\phi,\psi)$. Here we consider the case where two space orthogonal to all Killing vector fields $\xi_t$ and $\xi_{I}(I=\phi,\psi)$ is integrable. From the Frobenius conditions, $w^I_1=w^I_2=0$. Hence the metric can be written in Weyl-Papapetrou-type form:
\begin{eqnarray}
ds^2=f^{-1}e^{2\sigma}(d\rho^2+dz^2)-f^{-1}\rho^2dt^2+f_{IJ}(dx^I+w^I_3dt)(dx^J+w^J_3dt).
\end{eqnarray}
All the metric functions depend only on $\rho$ and $z$. The differential equations of the scalar fields are given by the axisymmetric solutions of Eqs.(\ref{eq:f}) and (\ref{eq:o}) on the abstract flat three surface with metric
\begin{eqnarray}
\tilde \gamma=d\rho^2+dz^2+\rho^2d\varphi^2,
\end{eqnarray}
which is written in the cylindrical coordinates. Namely, $D^2\tilde\phi$ and $D\tilde\phi\cdot D\tilde\psi$ are replaced with $\tilde\phi_{,\rho\rho}+\rho^{-1}\tilde\phi_{,\rho}+\tilde\phi_{,zz}$ and $\tilde\phi_{,\rho}\tilde\psi_{,\rho}+\tilde\phi_{,z}\tilde\psi_{,z}$, respectively.
 Once the five potentials $f_{IJ}$ and $\omega_I$ are obtained from Eqs.(\ref{eq:f}) and (\ref{eq:o}), Eq.(\ref{eq:gamma}) reduce to the equations with respect to the gradient of the metric function $\sigma$:
\begin{eqnarray}
{2 \over \rho}\sigma_{,\rho} &=&
{1 \over 4} f^{-2}[(f_{,\rho})^2-(f_{,z})^2]
+{1 \over 4} f^{IJ}f^{MN}(f_{IM,\rho}f_{JN,\rho}-f_{IM,z}f_{JN,z})
\nonumber\\&&
+{1 \over 2} f^{-1}f^{IJ}
(\omega_{I,\rho}\omega_{J,\rho}-\omega_{I,z}\omega_{J,z}),
\\
{1 \over \rho}\sigma_{,z} &=&
{1 \over 4} f^{-2} f_{,\rho} f_{,z}
+{1 \over 4} f^{IJ}f^{MN} f_{IM,\rho} f_{JN,z}
+{1 \over 2} f^{-1}f^{IJ} \omega_{I,\rho} \omega_{J,z},
\\
w^I_{3,\rho} &=&
\rho f^{-1} f^{IJ} \omega_{J,z},
\\
w^I_{3,z} &=&
-\rho f^{-1} f^{IJ} \omega_{J,\rho}.
\end{eqnarray}
This system is described by the following action which is invariant under the global $SL(3,{\bf R})$ transformation:
\begin{eqnarray}
S=\int d\rho dz\rho \left[\frac{1}{4}f^{-2}(\partial f)^2+\frac{1}{4}f^{IJ}f^{KL}\partial f_{IK}\cdot \partial f_{JK}+\frac{1}{2}f^{-1}f^{IJ}\partial\omega_I\cdot \partial \omega_J \right].\label{eq:action}
\end{eqnarray}
Here, we introduce the $SL(3,{\bf R})$ matrix $\Phi$ defined by
\begin{eqnarray}
\Phi=\left(
\begin{array}{ccc}
 f^{-1}&-f^{-1}\omega_\phi &-f^{-1}\omega_\psi \\
 -f^{-1}\omega_\phi&f_{\phi\phi}+f^{-1}\omega_\phi\omega_\phi &f_{\phi\psi}+f^{-1}\omega_\phi\omega_\psi \\
 -f^{-1}\omega_\psi &f_{\phi\psi}+f^{-1}\omega_\phi\omega_\psi & f_{\psi\psi}+f^{-1}\omega_\psi\omega_\psi 
\end{array}
\right),
\end{eqnarray}
where it is noted that this matrix is symmetric $({}^t\Phi=\Phi)$ and unimodular $({\rm det}\ \Phi=1 )$. 
Since we choose the Killing vector fields $\xi_\phi$ and $\xi_\psi$ to be spacelike, all the eigenvalues of $\Phi$ are real and positive. Therefore, there exists an $SL(3,{\bf R})$ matrix $g$ such that
\begin{eqnarray}
\Phi=g^tg.
\end{eqnarray} 
The square root matrix $g$ is determined up to the global $SO(3)$ transformation. In fact, under the rotation $g\to \Lambda g$ for any $\Lambda \in SO(3)$, $\Phi$ is invariant. Hence, the action describes a non-linear sigma model on the symmetric space $SL(3,{\bf R})/SO(3)$.
We define a current matrix as
\begin{eqnarray}
J_i=\Phi^{-1}\partial_i\Phi,
\end{eqnarray}
which is conserved if the scalar fields are the solutions of the equation of motion derived by the action (\ref{eq:action}). Then, the action (\ref{eq:action}) can be written in terms of $J$ and $\Phi$ as follows
\begin{eqnarray}
S&=&\frac{1}{4}\int d\rho dz \rho {\rm tr}(J_iJ^i)\\
 &=&\frac{1}{4}\int d\rho dz \rho {\rm tr}(\Phi^{-1}\partial_i\Phi\Phi^{-1}\partial^i\Phi).
\end{eqnarray}
Let us consider two sets of the field configuration $\Phi_{[0]}$ and $\Phi_{[1]}$ satisfying the equations Eq.(\ref{eq:f}) and Eq.(\ref{eq:o}). We denote the difference between the value of the functional obtained from the field configuration $\Phi_{[1]}$ and the value obtained from $\Phi_{[0]}$ as a bull's eye $\stackrel{\odot}{}$, e.g.,
\begin{eqnarray}
\stackrel{\odot}J{}^i=J^i_{[1]}-J^i_{[0]},
\end{eqnarray}
where the subscripts ${}_{[0]}$ and ${}_{[1]}$ denote the quantities associated with the field configurations $\Phi_{[0]}$ and $\Phi_{[1]}$, respectively.
The deviation matrix $\Psi$ is defined by
\begin{eqnarray}
\Psi=\stackrel{\odot}\Phi\Phi^{-1}_{[0]}=\Phi_{[1]}\Phi^{-1}_{[0]}-{\bf 1},
\end{eqnarray}
where ${\bf 1}$ is the unit matrix. 
Using the relation between the derivative of the deviation matrix and the $\stackrel{\odot}J{}^i$:
\begin{eqnarray}
D^i\Psi=\Phi_{[1]}\stackrel{\odot}J{}^i\Phi_{[0]}^{-1},\label{eq:deriv}
\end{eqnarray}
where $D$ is a covariant derivative associated with the abstract three-metric $\tilde \gamma$. 
The Mazur identity as the integration over the region $\Sigma=\{(\rho,z)|\rho\ge 0,\ -\infty<z<\infty \}$ is given by
\begin{eqnarray}
\int_{\partial \Sigma}\rho \partial_a{\rm tr}\Psi dS^a=\int_{\Sigma}\rho h_{ab}{\rm tr}({\cal M}^{a}{\ }^t{\cal M}^b)d\rho dz,\label{eq:id}
\end{eqnarray}
where $a,b$ run $\rho,z$, and $h=d\rho^2+dz^2$. 
The matrix ${\cal M}$ is defined by
\begin{eqnarray}
{\cal M}^a=g_{[0]}^{-1}{\ }^t\stackrel{\odot}J{}^ag_{[1]}.
\end{eqnarray}
It should be noted that the right hand side of the identity (\ref{eq:id}) is positive except the case of $\stackrel{\odot}J{}^i=0$. Therefore, we must have $\stackrel{\odot}J{}^i=0$ if the boundary conditions under which the left hand side of Eq.(\ref{eq:id}) vanishes are imposed at $\partial\Sigma$. Then, from Eq.(\ref{eq:deriv}), $\Psi$ is a constant matrix over the region $\Sigma$. To show that the limiting value of $\Psi$ is zero on at least one part of the boundary is sufficient to obtain the coincidence of two solutions $\Phi_{[0]}$ and $\Phi_{[1]}$.

\section{Boundary conditions and coincidence of solutions}\label{sec:coincidence}
The boundary integral in the left hand side of the Mazur identity (\ref{eq:id}) is decomposed into the integrals over the segments of the rod and the integral over the infinity as follows
\begin{eqnarray}
\int_{\partial \Sigma}\rho\partial_a{\rm tr}\Psi dS^a&=&\int_{-\infty}^{-ck^2}\rho\frac{\partial {\rm tr}\Psi }{\partial z}dz+\int_{-ck^2}^{ck^2}\rho\frac{\partial {\rm tr}\Psi }{\partial z}dz+\int_{ck^2}^{k^2}\rho\frac{\partial {\rm tr}\Psi }{\partial z}dz\nonumber\\
             &&+\int_{k^2}^{\infty}\rho\frac{\partial {\rm tr}\Psi }{\partial z}dz+\int_{\partial\Sigma_\infty}\rho\partial_a{\rm tr}\Psi dS^a,\label{eq:integral}
\end{eqnarray}
where $\partial\Sigma_\infty$ in the last term denotes the infinity. Now consider the line integrals of the twist one-forms $d\omega_{I}\ (I=\phi,\psi)$ over the $z$-axis in the canonical coordinates $(\rho,z)$. The twist one-forms vanish on the $\psi$-invariant planes and the $\phi$-invariant plane by its definition, i.e., the values of twist potentials are constant on the $z$-axis. Hence this can be expressed in the form
\begin{eqnarray}
\int^{ck^2}_{-ck^2}\omega_{I,z} dz=\biggl[\ \omega_I(z)\ \biggr]^{z=ck^2}_{z=-ck^2}\label{eq:line}
\end{eqnarray}
on the $z$-axis.
On the other hand, the left hand side of Eq.(\ref{eq:line}) can be written in terms of the Komar integral as follows
\begin{eqnarray}
\frac{1}{4\pi^2}\int_{{\cal H}}*d\xi_{I}=\frac{4}{\pi}J_I,
\end{eqnarray}
where ${\cal H}$ denotes the spatial cross section of the event horizon and $J_I(I=\phi,\psi)$ are angular momenta associated with the spacelike Killing vectors $\xi_I (I=\phi,\psi)$.
Using the ambiguity of the twist potentials $\omega_I$ in addition of a constant, without loss of generality, we always put their values on the $z$-axis such that
\begin{eqnarray}
\omega_I(z)=\frac{2J_I}{\pi}.\label{eq:twist}
\end{eqnarray}
for $z\in [ck^2,\infty]$, and
\begin{eqnarray}
\omega_I(z)=-\frac{2J_I}{\pi}\label{eq:twist2}
\end{eqnarray}
for $z\in [-\infty,-ck^2]$.
We estimate the integrals over the five boundaries and show that they vanish under the preferable boundary conditions.

(i) $\psi$-invariant planes: $\{(\rho,z)|\ \rho=0,\  -\infty<z\le -ck^2 \}$ and $\{(\rho,z)|\ \rho=0,\  ck^2\le z\le k^2\}$. The boundary integral over these regions are given by
\begin{eqnarray}
&&\int_{-\infty}^{-ck^2}\rho\frac{\partial {\rm tr}\Psi }{\partial z}\biggl|_{\rho=0}dz=\biggl[\rho{\rm tr}\Psi|_{\rho=0} \biggr]^{-ck^2}_{-\infty},\\
&& \int^{k^2}_{ck^2}\rho\frac{\partial {\rm tr}\Psi }{\partial z}\biggl|_{\rho=0}dz=\biggl[\rho{\rm tr}\Psi|_{\rho=0} \biggr]_{ck^2}^{k^2},
\end{eqnarray}
respectively.

We assume that for $\rho\to 0$, the five scalar fields behave as
\begin{eqnarray}
&&f_{\psi\psi[A]}\simeq f_{\psi\psi[A]}^{(2)}\rho^2+O(\rho^4),\label{eq:a1}\\
&&f_{\phi\phi[A]}\simeq f_{\phi\phi[A]}^{(0)}+O(\rho^2),\label{eq:a2}\\
&&f_{\phi\psi[A]}\simeq f_{\phi\psi[A]}^{(2)}\rho^2+O(\rho^4),\label{eq:a3}\\
&&\omega_{\phi[A]}\simeq \omega_{\phi[A]}^{(0)}+O(\rho^2),\label{eq:a4}\\
&&\omega_{\psi[A]}\simeq \omega_{\psi[A]}^{(0)}+O(\rho^2),\label{eq:a5}
\end{eqnarray}
where $A$ run $0,1$ and the coefficients $f_{IJ[A]}^{(2k)}$ and $\omega_{I[A]}^{(2k)}$ defined by $f_{IJ[A]}=\sum_{k=0}^{\infty}f_{IJ[A]}^{(2k)}\rho^{2k}$, $\omega_{I[A]}=\sum_{k=0}^\infty\omega_{I[A]}^{(2k)}\rho^{2k}$ are independent of $\rho$. The boundary condition (\ref{eq:a1}) comes from the requirement that $\rho=0$, $z\in[-\infty,-ck^2]$ and $\rho=0$, $z\in[ck^2,k^2]$ are the $\psi$-invariant plane, i.e., the plane invariant under the rotation with respect to the axial Killing vector $\partial_\psi$. The regularity on the invariant plane requires the other conditions (\ref{eq:a2})-(\ref{eq:a5}).
Hence, for $\rho\to 0$, $\rho {\rm tr}\Psi$ behaves as
\begin{eqnarray}
\rho {\rm tr}\Psi&\simeq& \frac{\left(\stackrel{\odot}{{\omega_{\psi}}^{(0)}}\right)^2}{f_{\psi\psi[0]}^{(2)}f_{\psi\psi[1]}^{(2)}f_{\psi\psi[0]}^{(2)}\rho^3}
+\frac{1}{f_{\psi\psi[0]}^{(2)2}f_{\psi\psi[1]}^{(2)2}f_{\phi\phi[0]}^{(0)2}f_{\phi\phi[1]}^{(0)}\rho}\nonumber\\
&&\times\Biggl[ \left(\stackrel{\odot}{{\omega_{\psi}}^{(0)}}\right)^2\left( -f_{\psi\psi[1]}^{(2)}f_{\phi\phi[1]}^{(0)}f_{[0]}^{(4)}-f_{\psi\psi[0]}^{(2)}f_{\phi\phi[0]}^{(0)}f_{[1]}^{(4)}-f_{\psi\psi[0]}^{(2)}f_{\psi\psi[1]}^{(2)}f_{\phi\phi[0]}^{(2)}f_{\phi\phi[1]}^{(0)}\right)\nonumber\\
&&+\left(\stackrel{\odot}{{\omega_{\phi}}^{(0)}}\right)^2f_{\psi\psi[0]}^{(2)}f_{\psi\psi[1]}^{(2)2}f_{\phi\phi[0]}^{(0)}+2\stackrel{\odot}\omega_\psi^{(0)}f_{\psi\psi[0]}^{(2)}f_{\psi\psi[1]}^{(2)}f_{\phi\phi[0]}^{(0)}(\stackrel{\odot}\omega_\psi^{(2)}f_{\phi\phi[1]}^{(0)}-\stackrel{\odot}\omega_\phi^{(0)}f_{\phi\psi[1]}^{(2)}) \Biggr]+O(\rho)\nonumber
\\
      &=&O(\rho),
\end{eqnarray}
where $f^{(4)}_{[A]}:=f_{\phi\phi[A]}^{(2)}f_{\psi\psi[A]}^{(2)}-f_{\phi\psi[A]}^{(2)2}$, and in the last equality we used the Eq.(\ref{eq:twist}) for the two solutions $\Phi_{[0]}$ and $\Phi_{[1]}$ with the same angular momenta.

(ii) $\phi$-invariant plane: $\{(\rho,z)|\ \rho=0,\ ck^2<z<\infty\}$. The boundary integral over the region is 
\begin{eqnarray}
\int^{\infty}_{ck^2}\rho\frac{\partial {\rm tr}\Psi }{\partial z}\biggl|_{\rho=0}dz=\biggl[\rho{\rm tr}\Psi|_{\rho=0} \biggr]_{ck^2}^{\infty}.
\end{eqnarray}
We assume that for $\rho\to 0$, the five scalar fields behave as
\begin{eqnarray}
&&f_{\phi\phi[A]}\simeq\rho^2f_{\phi\phi[A]}^{(2)}+O(\rho^4),\label{eq:b1}\\
&&f_{\psi\psi[A]}\simeq f_{\psi\psi[A]}^{(0)}+O(\rho^2),\label{eq:b2}\\
&&f_{\psi\phi[A]}\simeq\rho^2f_{\psi\phi[A]}^{(2)}+O(\rho^4),\label{eq:b3}\\
&&\omega_{\phi[A]}\simeq \omega_{\phi[A]}^{(0)}+O(\rho^2),\label{eq:b4}\\
&&\omega_{\psi[A]}\simeq \omega_{\psi[A]}^{(0)}+O(\rho^2),\label{eq:b5}
\end{eqnarray}
where the boundary condition (\ref{eq:b1}) comes from the requirement that $\rho=0$,$z\in[ck^2,\infty]$ is the $\phi$-invariant plane. The regularity on the invariant plane, i.e., the finiteness of the scalar fields, requires the other conditions (\ref{eq:b2})-(\ref{eq:b5}).
Hence, for $\rho\to 0$, $\rho {\rm tr}\Psi$ behaves as
\begin{eqnarray}
\rho {\rm tr}\Psi&\simeq& \frac{\left(\stackrel{\odot}{{\omega_{\phi}}^{(0)}}\right)^2}{f_{\phi\phi[0]}^{(2)}f_{\phi\phi[1]}^{(2)}f_{\phi\phi[0]}^{(2)}\rho^3}
+\frac{1}{f_{\phi\phi[0]}^{(2)2}f_{\phi\phi[1]}^{(2)2}f_{\psi\psi[0]}^{(0)2}f_{\psi\psi[1]}^{(0)}\rho}\nonumber\\
&&\times\Biggl[ \left(\stackrel{\odot}{{\omega_{\phi}}^{(0)}}\right)^2\left( -f_{\phi\phi[1]}^{(2)}f_{\psi\psi[1]}^{(0)}f_{[0]}^{(4)}-f_{\phi\phi[0]}^{(2)}f_{\psi\psi[0]}^{(0)}f_{[1]}^{(4)}-f_{\phi\phi[0]}^{(2)}f_{\phi\phi[1]}^{(2)}f_{\psi\psi[0]}^{(2)}f_{\psi\psi[1]}^{(0)}\right)\nonumber\\
&&+\left(\stackrel{\odot}{{\omega_{\psi}}^{(0)}}\right)^2f_{\phi\phi[0]}^{(2)}f_{\phi\phi[1]}^{(2)2}f_{\psi\psi[0]}^{(0)}+2\stackrel{\odot}\omega_\phi^{(0)}f_{\phi\phi[0]}^{(2)}f_{\phi\phi[1]}^{(2)}f_{\psi\psi[0]}^{(0)}(\stackrel{\odot}\omega_\phi^{(2)}f_{\psi\psi[1]}^{(0)}-\stackrel{\odot}\omega_\psi^{(0)}f_{\psi\phi[1]}^{(2)}) \Biggr]+O(\rho)\nonumber
\\
      &=&O(\rho),
\end{eqnarray}
where we also used Eq.(\ref{eq:twist}) for the two solutions $\Phi_{[0]}$ and $\Phi_{[1]}$ with the same angular momenta.

(iii) Horizon: $\{(\rho,z)|\ \rho=0$, \ $-ck^2<z<ck^2\}$. The regularity on the horizon requires the following behavior of the five-scalar fields for $\rho\to0$:
\begin{eqnarray}
&&f_{\phi\phi[A]}\simeq f_{\phi\phi[A]}^{(0)}+O(\rho^2),\\
&&f_{\psi\psi[A]}\simeq f_{\psi\psi[A]}^{(0)}+O(\rho^2),\\
&&f_{\psi\phi[A]}\simeq f_{\psi\phi[A]}^{(0)}+O(\rho^2),\\
&&\omega_{\phi[A]}\simeq \omega_{\phi[A]}^{(0)}+O(\rho^2),\\
&&\omega_{\psi[A]}\simeq \omega_{\psi[A]}^{(0)}+O(\rho^2).
\end{eqnarray}
Therefore, $\rho {\rm tr}\Psi$ on the horizon behaves as
\begin{eqnarray}
\rho{\rm tr}\Psi=O(\rho).
\end{eqnarray}

(iv) Infinity: $\partial\Sigma_\infty=\{(\rho,z)|\ \sqrt{\rho^2+z^2}\to\infty$ with $z/\sqrt{\rho^2+z^2}$ finite $\}$. From the analysis in Ref.~\cite{Harmark}, the asymptotic flatness requires that the metric in the canonical coordinates behaves as
\begin{eqnarray}
&&f_{\phi\phi[A]}\simeq (\sqrt{\rho^2+z^2}-z)\left(1+\frac{\tilde f_{\phi\phi[A]}^{(2)}}{2\sqrt{\rho^2+z^2}}\right)+O\left(\frac{1}{\rho^2+z^2}\right),\\
&&f_{\psi\psi[A]}\simeq (\sqrt{\rho^2+z^2}+z)\left(1+\frac{\tilde f_{\psi\psi[A]}^{(2)}}{2\sqrt{\rho^2+z^2}}\right)+O\left(\frac{1}{\rho^2+z^2}\right),\\
&&f_{\psi\phi[A]}\simeq \zeta\frac{\rho^2}{(\rho^2+z^2)^{3/2}}+O\left(\frac{1}{\rho^2+z^2}\right),\\
&& \omega_{\phi[A]}\simeq \tilde \omega_{\phi[A]}^{(0)}+O\left(\frac{1}{\sqrt{\rho^2+z^2}}\right),\\
&&\omega_{\psi[A]}\simeq \tilde \omega_{\psi[A]}^{(0)}+O\left(\frac{1}{\sqrt{\rho^2+z^2}}\right)
\end{eqnarray}
for $\sqrt{\rho^2+z^2}\to\infty$ with $z/\sqrt{\rho^2+z^2}$ finite, where $\zeta$ is a gauge-invariant constant. For two solutions with the same masses $M$, the functions $\tilde f_{\phi\phi[A]}^{(2)}$ and $\tilde f_{\psi\psi[A]}^{(2)}$ can be written in the form
\begin{eqnarray}
\tilde f_{\phi\phi[A]}^{(2)}=\frac{4}{3\pi}(M+\eta),\quad \tilde f_{\psi\psi[A]}^{(2)}=\frac{4}{3\pi}(M-\eta),
\end{eqnarray}
where the constant $\eta$ is not gauge-invariant. The ambiguity in the choice of the parameter $\eta$ means that the coordinate $z$ is uniquely determined up to the translation $z\to z+\alpha$ as far as the coordinate $\rho$ conjugate to $z$  is fixed at the infinity. Since in our proof we choose the coordinate $z$ such that the horizons are located on $[-ck^2,ck^2]$ for two configurations $\Phi_{[0]}$ and $\Phi_{[1]}$, we choose the same values of $\eta$ for the two solutions. 
Under the choice of this gauge, $\rho {\rm tr}\Psi$ behaves as
\begin{eqnarray}
\rho{\rm tr}\Psi&\simeq& \frac{1}{4}\left[\left(\stackrel{\odot}{\tilde f_{\phi\phi}^{(2)}}\right)^2+\left(\stackrel{\odot}{\tilde f_{\psi\psi}^{(2)}}\right)^2+\stackrel{\odot}{\tilde f_{\phi\phi}^{(2)}}\stackrel{\odot}{\tilde f_{\psi\psi}^{(2)}}\right]\frac{\rho}{\rho^2+z^2}+O\left(\frac{1}{\rho^2+z^2}\right)\\                 
                 &=&O\left(\frac{1}{\rho^2+z^2}\right).
\end{eqnarray}
in the neighborhood of the infinity. Hence in the neighborhood of $\partial\Sigma_\infty$,
\begin{eqnarray}
\rho\partial_a{\rm tr\Psi}dS^a=O\left(\frac{1}{\sqrt{\rho^2+z^2}}\right).
\end{eqnarray}

From (i)-(iv), the boundary integral (\ref{eq:integral}) vanishes on each segment of the rod and the infinity. The deviation matrix $\Psi$ is constant and has asymptotic behavior as $\Psi\to 0$. Hence $\Psi$ vanishes over $\partial\Sigma$. Thus the two configurations $\Phi_{[0]}$ and $\Phi_{[1]}$ coincides with each other.

\section{Conclusions and Discussions}\label{sec:conclusion}
\newtheorem{pro}{Proposition}
\newtheorem{thm}{Theorem}
\newtheorem{co}{Corollary}
The rigidity theorem of the four-dimensional black hole space-times states that asymptotically flat, stationary, analytic spacetimes assure the existence of a axial Killing vector field~\cite{Hawking2}. Recently, the rigidity theorem was generalized to higher-dimensional space-times~\cite{Ishibashi}. In particular, the rigidity theorem in five-dimensions guarantees at least one axial Killing vector field. As conjectured by Reall~\cite{Reall}, there may exist black hole solutions admitting only two commuting Killing vector fields, although all of five-dimensional stationary black hole solutions found so far have three mutually commuting Killing vector fields. Hence, as this stage, it is natural to concentrate on asymptotically flat black hole solutions to the five-dimensional vacuum Einstein equations admitting three commuting Killing vector fields. 
One of authors and Ida~\cite{Morisawa:2004tc} showed that the only black hole solution with a regular event horizon homeomorphic to $\rm S^3$ is the five-dimensional Myers-Perry black hole solution in this class. However, it is impossible to generalize this theorem to the solutions with horizon topology $\rm \rm S^1\times \rm S^2$ since there are two black ring solutions with different shapes for the same mass and the same angular momenta~\cite{Emparan:2001wn,Pom}.
 Hence, we must introduce some additional geometrical information in order to consider the uniqueness of black rings as the boundary value problem. For instance, one of the candidates is rod structure introduced by Harmark~\cite{Harmark}.  As mentioned in Sec.\ref{sec:boundary}, the black ring solution have the following rod structure:
\begin{eqnarray}
&&(i) [-\infty,z_1], \quad v=(0,0,1),\nonumber\\
&&(ii) [z_1,z_2], \quad v=(1,\Omega_1,\Omega_2)\nonumber\\
&&(iii) [z_2,z_3], \quad v=(0,0,1),\nonumber\\ 
&&(iv) [z_3,\infty],\quad v=(0,1,0),\label{eq:rod}
\end{eqnarray}
which is characterized by four segments $[z_i,z_{i+1}]\ (i=0,1,\cdots,3; z_0=-\infty,z_4=\infty)$ and eigenvectors $v$ with respect to a zero eigenvalue of the three-dimensional matrix $g_{\alpha\beta}$ for each segments. The constants $\Omega_1$ and $\Omega_2$ mean the angular velocities of the horizon in the directions of $\partial_\phi$ and $\partial_\psi$, respectively.

  By introducing the rod structure~\cite{Harmark},
Hollands and Yazadjiev applied the discussion in Ref.~\cite{Morisawa:2004tc} to the case of non-spherical horizon topology and showed the following theorem in Ref.\cite{Hollands}.
\begin{thm}\label{eq:thm}
Consider two stationary, asymptotically flat, vacuum black objects spacetimes of the five-dimensions with commuting two axial Killing vector fields and a timelike Killing vector field. Then, if both solutions have the same topology, the same rod structure and the values of the mass $M$ and angular momenta $J_1$, $J_2$, they are isometric.
\end{thm}

However, even if we restrict the horizon topology to $\rm \rm S^1\times \rm S^2$, this theorem does not imply the uniqueness of the Pomeransky-Sen'kov black ring solution within the class of these solutions since there may exist another black ring solution admitting different rod structures without a conical singularity for the same asymptotic charges, mass and two angular momenta.  
Our end of this article is to prove the uniqueness of the Pomeransky-Sen'kov black ring solution by showing that the black rings with the rod structures different from that of the Pomeransky-Sen'kov black ring solution have conical singularities.
Our proof is composed of two steps: First, we show the existence of asymptotically flat black ring solution with conical singularities and without curvature singularities such that under the condition of no conical singularity, they coincides with the Pomeransky-Sen'kov black ring solution; next, once these black ring solution is given, using the theorem \ref{eq:thm} obtained by Hollands and Yazadjiev, we can show the uniqueness of the black ring solution (\ref{eq:metric}) in this class of the solutions admitting three mutually commuting Killing vector fields, i.e., a timelike Killing vector field and two axial Killing vector fields. However, we should note the following point. If we apply the Hollands-Yazadjiev's theorem to this black ring solution (\ref{eq:metric}), it seems to be specified by the asymptotic charges $M,J_\phi,J_\psi$ and the four additional parameters $c,k,\Omega_1,\Omega_2$ appearing in the rod data, although all of these parameters are not independent. In the proof in Sec.\ref{sec:coincidence} the only four parameters $M,J_\phi,J_\psi$ and $c$ appear, where $M,J_I(I=\phi,\psi)$ denote the mass and angular
  momenta, respectively, and the constant $c$ has the geometrical meaning of the ratio of the radius of $\rm S^2$ to the radius of $\rm S^1$. In terms of these parameters, we obtain the following result:
\begin{co}
Consider asymptotically flat black ring solutions to the five-dimensional vaccum Einstein equations admitting three commuting Killing vector fields, i.e., two axial Killing vector fields and a timelike Killing vector field. Then, in this class of solutions, the only solution with the horizon topology of $S^1\times S^2$ is the black ring solution (\ref{eq:metric}) specified by a mass $M$, two angular momenta $J_\phi,J_\psi$ and the ratio $c$ of the radius of $S^2$ to the radius of $ S^1$.
\end{co}
In particular, if we impose that the black ring solutions do not admit a conical singularity, we obtain the main result in this article:

\begin{thm}
The only asymptotically flat, five-dimensional black ring solution with commuting two axial Killing vector fields and a timelike Killing vector field and without a conical singularity is the Pomeransky-Sen'kov solution. 
\end{thm}

\section*{Acknowledgements}
We thank Daisuke Ida  and Hideo Kodama for useful comments. We also thank A. A. Pomeransky and R. A. Sen'kov for informing us of the seed solution generating the black ring solution with two angular momenta. The work of YY is supported by the Grant-in Aid for Scientific Research (No. 19540304 and No. 19540098) from Japan Ministry of Education.

\appendix
\section{explicit expressions of metric functions}
\label{app:metric}
The one-form $\Omega$ appeared in the general black ring 
metric~(\ref{eq:metric}) is defined by 
\begin{eqnarray}
\Omega=\Omega_{\phi}d\phi+\Omega_{\psi}d\psi,\nonumber
\end{eqnarray}
where
\begin{eqnarray}
& &\Omega_{\phi}=
\frac{{\sq{2}}^{2} \sq{3} {k}}{32 {\sq{1}}^{2} {\beta} H(y,x)} 
{\sqrt {{\frac {b \left( 1-b \right) \left( b\SQ+4 q \right) {\SQ}}{2 {\cY}^{3}}}}}
\times
\nonumber\\&&
\left\{ \left[ 8 b{q}^{3}{\SQ}^{2}{\sq{1}}^{2} \left( b-1 \right)  \left( 1+b \right)  \left( {\sq{1}}^{2}\Sx  \left( y-1 \right) ^{2}-4 \Ty \Ap \right)  \left( x-1 \right) {\alphazero}^{2}
\right.\right.
\nonumber\\&&
\quad
\left.\left.
-4 {b}^{2}{q}^{2}\SQ {\sq{1}}^{4}\sq{3}  \left( \sq{3} \Sx {\sq{2}}^{2} \left( y-1 \right) ^{2} \left( 1+b \right) -2 \Cpx {\sq{2}}^{2} \left( y-1 \right) ^{2}+4 \sq{0} \Sy \Am  \left( b-1 \right)  \right)  \left( x-1 \right) \alphazero \right] {\beta}^{2}
\right.
\nonumber\\&&
\left.
+ \left[ 8 \Ty {\Tx}^{2}{\SQ}^{3}{q}^{3} \left( b-1 \right) \left( 1+b \right) ^{2} \left( 1+y \right) {\alphazero}^{3}
-4 b\sq{3} {q}^{2}{\SQ}^{2} \left( 1+b \right) 
\times
\right.\right.
\nonumber\\&&
\quad
\left.\left.\left( 8 \Ty q\Am \sq{1}  \left( b-1 \right)  \left( x+1 \right) +2 \Bpy {\Tx}^{2}\Ty {\sq{2}}^{2}
-2 \Ap \Am {\sq{1}}^{2} \left( b-1 \right)+\sq{3} \Ty {\Tx}^{2}{\sq{2}}^{2} \left( 1+b \right)  \left( 1+y \right)  \right) {\alphazero}^{2}
\right.\right.
\nonumber\\&&
\quad
\left.\left.
+2 {b}^{2}q\SQ {\sq{1}}^{2}\sq{2} {\sq{3}}^{3} \left( 1+b \right)  \left( -{\sq{1}}^{2}\Sy {\Sx}^{2} \left( y-1 \right) +2 \Ap \Am \sq{2}+8 \Ap \Sy q \left( x+1 \right)  \right) \alphazero
\right.\right.
\nonumber\\&&
\quad
\left.\left.
+{b}^{3}{\Sx}^{2}{\sq{1}}^{4}\Sy {\sq{3}}^{5}{\sq{2}}^{2}\sq{0}  \left( 1+y \right)  \right] \beta
\right.
\nonumber\\&&
\left.
-8 {\sq{3}}^{3}b{q}^{3}\sq{2} {\SQ}^{2}\Tx  \left( 1+b \right) ^{2} \left( 1+y \right) ^{2} \left( x+1 \right) {\alphazero}^{2}-4 {b}^{2}{\sq{3}}^{5}{q}^{2}{\sq{2}}^{2}\SQ \Sx \sq{1}  \left( 1+y \right) ^{2} \left( 1+b \right)  \left( x+1 \right) \alphazero \right\} \nonumber
,\\
& &\Omega_{\psi}=
-\frac{b \SQ \left( 1+b \right) ^{2} {q}^{2} {\sq{2}}^{2}\sq{3} k \left( 1-x \right)  \left( 1+x \right)}
{4 \sq{1} {H(y,x)}}
\sqrt{\frac {\alphazero}{\left( 1+b \right) {\cY}^{3}\beta}}
\times
\nonumber\\&&
\frac
{\left( 4 {q}^{2}{\SQ}^{2} \left( b-1 \right)  \left( 1+b \right) {\alphazero}^{2}+{b}^{2}{\sq{2}}^{2}{\sq{0}}^{2}{\sq{1}}^{2}{\sq{3}}^{2} \right) }
{\left( 2  \left( 1+b \right) q \SQ \alphazero+b{\sq{0}}^{2}{\sq{1}}^{2} \right) ^{2}}
\left( 4 {q}^{2}{\SQ}^{2} \left( b-1 \right)  \left( 1+b \right)  \left( -q+2 y+\lambda \right) \Ty {\alphazero}^{2}
\right.
\nonumber\\&&
\left.
+4 bq\sq{1} \sq{3} \SQ  \left( y{\lambda}^{2}+2 \lambda+2 {y}^{2}\lambda+4 y-{q}^{2}y \right)  \left( \SQ+4 bq \right) \alphazero
+{b}^{2}\sq{0} \sq{2} {\sq{3}}^{3}{\sq{1}}^{3} \left( \lambda+q+2 y \right) \Sy \right)\nonumber
.
\end{eqnarray}
The functions $H,J,F$ are defined as:
\begin{eqnarray}
H(x,y)&=&
\frac {\SQ {\sq{2}}^{2}}{64 \cY {\sq{1}}^{2} \beta}
\left\{ 4 \Cpy b{q}^{2}\SQ {\sq{1}}^{4} \left( x-1 \right) ^{2} \left( b-1 \right)  \left( y-1 \right) \alphazero {\beta}^{2}
\right.
\nonumber\\&&
\quad
\left.
+ \left[ -4 {\Ty}^{2}\Bpx {q}^{2}\Tx {\SQ}^{2} \left( b-1 \right)  \left( 1+b \right) {\alphazero}^{2}
+4 qb\Ap \Am \SQ {\sq{1}}^{2}{\sq{3}}^{2} \left( b-1 \right)  \left( 1+b \right) \alphazero
\right.\right.
\nonumber\\&&
\qquad
\left.\left.
+{b}^{2}{\Sy}^{2}\Sx \Bmx {\sq{1}}^{4}{\sq{3}}^{4} \right] \beta
-4 \Cmy b{q}^{2}\SQ {\sq{3}}^{4} \left( x+1 \right) ^{2} \left( 1+b \right)  \left( 1+y \right) \alphazero \right\},\nonumber
\\
J(x,y)&=&
\frac{-1}{8 \left( x-y \right) {\sq{1}}^{2}{\cY}^{2} k}
\sqrt{
\frac {\SQ \left( 1-b \right) }{2 b \left( b\SQ+4 q \right)  \left( 1+b \right) \alphazero \beta}
}
\left\{ \left( b\SQ+4 q \right) {k}^{3} \left( 1+b \right) \alphazero q\sq{3}  \left( 1+y \right)  \left( x+1 \right)  
\right.
\nonumber\\&&
\left.
\left( 8 b{q}^{2}\SQ {\sq{1}}^{2}\alphazero  \left( b-1 \right)  \left( y-1 \right)  \left( x-1 \right)  \left( 2 q \left( b-1 \right) \SQ  \left( \sq{3}  
\Dxy
\left( 1+b \right) -\sq{1} {\sq{0}}^{2}\Ty \Tx \right) \alphazero
\right.\right.\right.
\nonumber\\&&
\left.\left.\left.
-\sq{3} b{\sq{1}}^{2}\sq{0}  \left( \sq{0}  
\Dxy
\left( b-1 \right) +\sq{2} {\sq{3}}^{2}\Sy \Sx \right)  \right) {\beta}^{2}
\right.\right.
\nonumber\\&&
\left.\left.
+ \left( -8 {q}^{3}\Ty \Tx {\SQ}^{3} \left( 1+b \right)  \left( b-1 \right) ^{2} \left( \sq{3} \Am  \left( 1+b \right) -4 q \left( y-1 \right)  \left( x-1 \right) \sq{1} \right) {\alphazero}^{3}
\right.\right.\right.
\nonumber\\&&
\left.\left.\left.
-4 {q}^{2}{\sq{3}}^{3}\Sx {\SQ}^{2}b\sq{1} \Sy  \left( b-1 \right)  \left( 1+b \right)  \left( \sq{1} \Ap  \left( b-1 \right) -2 \sq{2} \Ty \Tx \right) {\alphazero}^{2}
\right.\right.\right.
\nonumber\\&&
\left.\left.\left.
+2 q \left( b-1 \right) \Ty {\sq{3}}^{3}{b}^{2}\SQ {\sq{1}}^{2}\sq{2} {\sq{0}}^{2}\Tx  \left( \sq{2} \Am  \left( 1+b \right) +2 \sq{1} \Sy \Sx \right) \alphazero
\right.\right.\right.
\nonumber\\&&
\left.\left.\left.
+{\sq{3}}^{5}\Sx {b}^{3}{\sq{2}}^{2}{\sq{1}}^{4}\Sy \sq{0}  \left( \sq{0} \Ap  \left( b-1 \right) -4 q \left( y-1 \right)  \left( x-1 \right) \sq{2} \right)  \right) \beta
\right.\right.
\nonumber\\&&
\left.\left.
+ \left( 1+b \right) \sq{3}  \left( 2 q\SQ  \left( b-1 \right) \alphazero-b{\sq{2}}^{2}{\sq{3}}^{2} \right)  
\left( 4 \Ty {q}^{2}\Am {\SQ}^{2}\Tx  \left( b-1 \right)  \left( 1+b \right) {\alphazero}^{2}
\right.\right.\right.
\nonumber\\&&
\left.\left.\left.
-4 bq \SQ \sq{1} \sq{2} {\sq{3}}^{2} \Sx \Sy \Ty \Tx \alphazero 
-{b}^{2} {\sq{1}}^{2}{\sq{2}}^{2}{\sq{3}}^{4} \Sx \Sy \Ap 
\right)  \right) {\sq{2}}^{2}b \right\}\nonumber
\end{eqnarray}
\begin{eqnarray}
F(x,y)&=&
\frac{-\left( 1+b \right) {b}^{3}{k}^{2}\Am \Ap  \left( b\SQ+4 q \right) {\sq{3}}^{2}\Sy {\sq{2}}^{2}}
{8 \Sx \Bpx \Ty {\cY}^{2}}
\times
\nonumber\\&&
\left( 8 \alphazero {q}^{2}\SQ \sq{1}  \left( b-1 \right)  \left( x-1 \right) \beta
+\sq{3}  \left( 2 \SQ q \left( b-1 \right)  \left( -4 q \left( x+1 \right) +\Sx \sq{1} \right) \alphazero
-\sq{1} \Sx b{\sq{2}}^{2}{\sq{3}}^{2} \right)  \right) ^{2} 
\nonumber\\&&
+
\frac {{k}^{2} \left( 1+b \right) \Tx \Ap {\sq{2}}^{2}b}
{16 {\Ty}^{2}{\Bpx}^{2} \left( x-y \right) {\Sx}^{2}{\sq{1}}^{2}{\cY}^{2}}
\times
\nonumber\\&&
\left\{ -16 {b}^{2}{\sq{3}}^{2}{q}^{2}\SQ {\Sy}^{2}{\sq{1}}^{4}\Sx  \left( x-1 \right)  \left( x+1 \right)  \left( b-1 \right)  \left( b\SQ+4 q \right) \alphazero \beta
\right.
\nonumber\\&&
\left.
+\Bpx  \left[ 4 {\SQ}^{2} \left( b-1 \right) {q}^{2} \left( {\Sx}^{2}{\sq{1}}^{4} \left( b-1 \right)  \left( y-1 \right) \Cpy-32 {q}^{2}\Ty  \left( x+1 \right) ^{2} \left( b\SQ+4 q \right) \Bpy
\right.\right.\right.
\nonumber\\&&
\left.\left.\left.
+8 {\sq{1}}^{2}q\Ty  \left( b-1 \right)  \left( y-1 \right)  \left( x+1 \right)  \left( b\SQ+4 q \right) \Sx \right) {\alphazero}^{2}
\right.\right.
\nonumber\\&&
\left.\left.
-4 b{\sq{3}}^{2}q\Sx \SQ {\sq{1}}^{2}{\sq{2}}^{2} \left( b-1 \right)  \left( y-1 \right)  \left( 4 q \left( b\SQ+4 q \right)  \left( x+1 \right) \Ty+{\sq{1}}^{2}\Cpy \Sx \right) \alphazero
\right.\right.
\nonumber\\&&
\left.\left.
+{\Sx}^{2}{\sq{1}}^{4}{b}^{2}{\sq{2}}^{4}{\sq{3}}^{4}\Cpy  \left( y-1 \right)  \right]  \right\} 
\nonumber\\&&
\times
\left( 4 {q}^{2} \SQ  \left( y-1 \right) \Ty  \left( b-1 \right)  \left( x-1 \right) \Bpx \alphazero \beta 
+b{\sq{3}}^{4} \left( x+1 \right) \Sx \Sy \Cmy \right)
\nonumber\\&&
-\frac{b{\sq{2}}^{2}\Am \Tx {k}^{2}}
{16{\cY}^{2}\beta {\sq{1}}^{2}{\Sx}^{2} \left( x-y \right) {\Bpx}^{2}{\Ty}^{2}}
\times
\nonumber\\&&
\left( -b{\sq{1}}^{4} \left( x-1 \right) \Sx \Sy \Cpy \beta+4 {q}^{2}\SQ  \left( 1+y \right) \Ty  \left( 1+b \right)  \left( x+1 \right) \Bpx \alphazero \right)  
\times
\nonumber\\&&
\left\{ -128 \Bpx \Bpy {q}^{4}{\alphazero}^{2}{\SQ}^{2}\Ty  \left( b-1 \right) ^{2} \left( x-1 \right) ^{2} \left( b\SQ+4 q \right) {\beta}^{2}
\right.
\nonumber\\&&
\left.
+16 \SQ {\sq{3}}^{2}{q}^{2}\Sx \alphazero  \left( b-1 \right)  \left( x-1 \right)  \left( b\SQ+4 q \right)  
\left[ 2 q\SQ \Ty \Bpx  \left( b-1 \right)  \left( 1+b \right)  \left( 1+y \right) \alphazero
\right.\right.
\nonumber\\&&
\left.\left.
-{\sq{3}}^{2}b \left( {\sq{2}}^{2}\Bpx  \left( 1+b \right)  \left( 1+y \right) \Ty+b{\Sy}^{2}{\sq{1}}^{2} \left( b-1 \right)  \left( x+1 \right)  \right)  \right] \beta
\right.
\nonumber\\&&
\left.
-{\sq{3}}^{4}\Bpx \Cmy {\Sx}^{2} \left( 1+b \right)  \left( 2 q\SQ  \left( b-1 \right) \alphazero-b{\sq{2}}^{2}{\sq{3}}^{2} \right) ^{2} \left( 1+y \right)  \right\} 
\nonumber\\&&
+\frac{{k}^{2} \left( 1+b \right) \Am \Ap \Tx q {\sq{2}}^{2}b \left( b-1 \right) }
{4 \left( x-y \right) {\Bpx}^{2}{\Ty}^{2}{\Sx}^{2}{\sq{1}}^{2}{\cY}^{2}}
\left( 2 \alphazero q\SQ \Ty \Bpx-b{\sq{1}}^{2}{\sq{3}}^{2}\Sy \Sx \right) 
\times
\nonumber\\&&
\left\{ 4 \SQ \alphazero q \left( b-1 \right)  \left( x-1 \right)  \left( b\SQ+4 q \right)  \left[ 2 \SQ q\Ty \Bpx  
\left( -8 \Bpy q \left( x+1 \right) +{\sq{1}}^{2} \left( b-1 \right)  \left( y-1 \right) \Sx \right) \alphazero
\right.\right.
\nonumber\\&&
\quad
\left.\left.
-b\Sx {\sq{3}}^{2}{\sq{1}}^{2} \left( {\sq{2}}^{2}\Ty  \left( y-1 \right) \Bpx+2 {\Sy}^{2}\sq{1} \sq{3} b \left( x+1 \right)  \right)  \right] \beta
\right.
\nonumber\\&&
\left.
+{\sq{3}}^{2}\Bpx \Sx  \left( 1+b \right)  \left( 1+y \right)  \left( 2 q\SQ  \left( b-1 \right) \alphazero-b{\sq{2}}^{2}{\sq{3}}^{2} \right)  
\right.
\times
\nonumber\\&&
\quad
\left.
\left[ 2 q\SQ  \left( 2  \left( b\SQ+4 q \right)  \left( x+1 \right) \Ty+{\sq{1}}^{2} \left( b-1 \right)  \left( y-1 \right) \Sx \right) \alphazero-b\Sx {\sq{1}}^{2}{\sq{2}}^{2}{\sq{3}}^{2} \left( y-1 \right)  \right]  \right\} 
\nonumber\\&&
-\frac{2 {k}^{2}\Sy {\Tx}^{2} {\it H(x,y)}}{\SQ \cY  \left( x-y \right) ^{2}\Ty \Bpx \Sx}
\times
\nonumber\\&&
\left\{ 32 {q}^{2}\left( b\SQ+4 q \right) \Bpy  \left( x-1 \right)  \left( x+1 \right) \Sy b\SQ  \left( b-1 \right) \alphazero \beta  
\right.
\nonumber\\&&
\quad
\left.
+\Bpx \Sx  \left( y-1 \right)  \left( 1+y \right)  \left( 1+b \right)  \left( 2 q\SQ  \left( b-1 \right) \alphazero-b{\sq{2}}^{2}{\sq{3}}^{2} \right) ^{2} \right\}, \nonumber
\end{eqnarray}
where the several polynomials are defined by
\begin{eqnarray}
\Ap &=& {q}^{2}xy+2 q \left( -1+xy \right) - \left( 2+x\lambda \right)  \left( 2+y\lambda \right), 
\\
\Am &=& {q}^{2}xy+2 q \left( 1-xy \right) - \left( 2+x\lambda \right)  \left( 2+y\lambda \right), 
\\
\Bpx &=& 2-bq+2 bx-qx+b\lambda+x\lambda,
\\
\Bmx &=& -2+bq+2 bx-qx+b\lambda-x\lambda,
\\
\Bpy &=& 2-bq+2 by-qy+b\lambda+y\lambda,
\\
\Bmy &=& -2+bq+2 by-qy+b\lambda-y\lambda,
\\
\Cpx &=& -4 b+2 q-2 qx+b{q}^{2}x-2 b\lambda-2 bx\lambda-bx{\lambda}^{2},
\\
\Cmx &=& 4 b-2 q-2 qx+b{q}^{2}x-2 b\lambda+2 bx\lambda-bx{\lambda}^{2},
\\
\Cpy &=& -4 b+2 q-2 qy+b{q}^{2}y-2 b\lambda-2 by\lambda-by{\lambda}^{2},
\\
\Cmy &=& 4 b-2 q-2 qy+b{q}^{2}y-2 b\lambda+2 by\lambda-by{\lambda}^{2},
\\
\Dxy &=& {q}^{4}xy- ( 4 xy+2 {\lambda}^{2}xy+2 \lambda x+2 \lambda y-4 ) {q}^{2}- ( 2+\lambda )  ( 2-\lambda )  ( 2+x\lambda )  ( 2+y\lambda ),
\\
\Sx &=& 2+x\lambda+qx,
\\
\Sy &=& 2+y\lambda+qy,
\\
\Tx &=& -2-x\lambda+qx,
\\
\Ty &=& -2-y\lambda+qy,
\end{eqnarray}
and the constants are defined by
\begin{eqnarray}
\SQ &=& -4+{\lambda}^{2}-{q}^{2},
\\
\sq{0} &=& 2+q+\lambda,
\\
\sq{1} &=& 2+q-\lambda,
\\
\sq{2} &=& -2+q+\lambda,
\\
\sq{3} &=& -2+q-\lambda,
\\
\cY &=& 32 b{q}^{2}\SQ  ( b-1 )  ( b\SQ+4 q ) \alphazero \beta
+ ( 1+b )  ( 2 ( b-1 ) q\SQ \alphazero  -b{\sq2}^{2}{\sq3}^{2} ) ^{2},
\\
\beta &=& {\frac 
{ ( 1+b )  ( 2  ( 1-b ) q \SQ \alphazero+b{\sq2}^{2}{\sq3}^{2} ) }
{ ( 1-b )  ( 2  ( 1+b ) q \SQ \alphazero+b{\sq0}^{2}{\sq1}^{2} ) }}.
\end{eqnarray}
Apparently, here are seven parameters $\lambda,\nu,q,c,\alphazero,b,k$.
These parameters obey the following three relations:
\begin{eqnarray}
q &=& \sqrt{\lambda^2-4\nu},
\\
c &=& q/(1-\nu),
\\
\alphazero &=& 4 \frac 
{ ( \lambda+q ) ( \lambda+2-q ) ( \lambda-2-q ) }
{ ( \lambda-q ) ( \lambda+2+q ) ( \lambda-2+q ) }.
\end{eqnarray}
Thus the four of them are independent.

\end{document}